\documentclass[mathleft
]{an}
\usepackage{graphicx}
\usepackage{times}
\usepackage{amsmath}
\usepackage{amssymb}
\usepackage{url,hyperref}

\begin{document}

\Pagespan{1}{}
\Yearsubmission{2015}%

\title{Modelling aperiodic X-ray variability in black hole binaries as
  propagating mass accretion rate fluctuations: a short review}

\author{Adam Ingram\inst{1}\fnmsep\thanks{Corresponding author:
  \email{a.r.ingram@uva.nl}\newline}
}
\titlerunning{Propagating fluctuations}
\authorrunning{A. R. Ingram}
\institute{
Anton Pannekoek Institute
University of Amsterdam,
Science Park 904, 1098 XH Amsterdam, the Netherlands
}

\received{In prep}
\publonline{later}


\abstract{
Black hole binary systems can emit very bright and rapidly varying
X-ray signals when material from the companion accretes onto the black
hole, liberating huge amounts of gravitational potential
energy. Central to this process of accretion is turbulence. In the
propagating mass accretion rate fluctuations model, turbulence is
generated throughout the inner accretion flow, causing fluctuations in
the accretion rate. Fluctuations from the outer regions propagate
towards the black hole, modulating the fluctuations generated in the
inner regions. Here, I present the theoretical motivation behind this
picture before reviewing the array of statistical variability
properties observed in the light curves of black hole binaries that
are naturally explained by the model. I also discuss the remaining
challenges for the model, both in terms of comparison to data and in
terms of including more sophisticated theoretical considerations.
}

\maketitle

\section{Introduction}

\begin{figure}
\centering
 \includegraphics[height= 75mm,width=80mm,trim=6.5cm 2.0cm 5.8cm
2.5cm,clip=true]{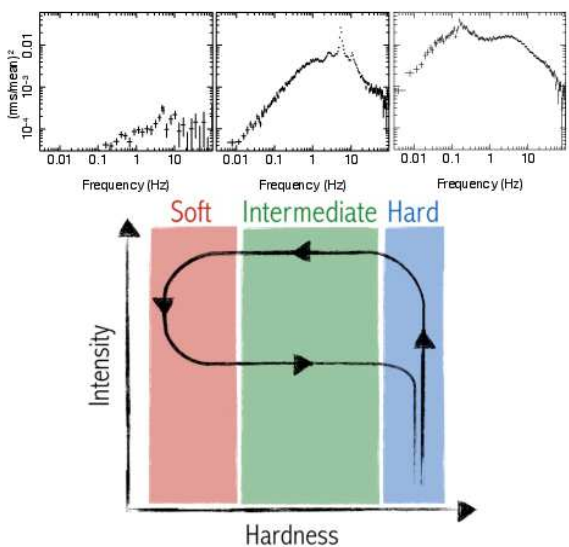}
 \caption{
\textit{Bottom:} Schematic demonstrating the path traced on a
hardness intensity diagram (HID) for a typical (idealised) black hole
binary outburst. The source passes through three broadly defined
states, the hard (blue), intermediate (green) and soft (red)
states. \textit{Top:} Power spectra representative of these three
states, with the hard and soft states on the right and left
respectively. Here, the power spectra are plotted as frequency
$\times$ power, such that the y-axis is in units a fractional rms
variability.}
\label{fig:HIDschem}
\vspace{-5mm}
\end{figure}

Black hole binaries display variability in their X-ray light curves on
a wide range of timescales. On the longest timescales ($\sim$weeks to
months), their spectrum evolves dramatically from being dominated by a
hard power law in the \textit{hard state}, to being dominated by a
quasi-thermal accretion disk in the \textit{soft state} (Tananbaum et
al 1972; Done, Gierlinski \& Kubota 2007). These spectral changes can
be illustrated using a \textit{hardness intensity diagram} (HID: Homan
et al 2001; Belloni et al 2005). The x-axis is hardness, typically
defined as the ratio of photon counts in the $16-20$ keV and $2-6$ keV
bands of the \textit{Rossi X-ray Timing Explorer} (\textit{RXTE}), and
the y-axis is intensity, typically defined as counts in the full
\textit{RXTE} band. This has the advantage of being completely model
independent and it also allows a large amount of information to be
condensed into a single plot. That the 
definitions are rather married to \textit{RXTE} is a symptom of just
how large a fraction of the available data is from that particular
mission. Fig \ref{fig:HIDschem} (bottom) shows schematically the 
tracks followed by a typical transient black hole binary
outburst. Starting at the bottom right-hand corner in quiescence, the
source is initially in the hard state as its luminosity starts to
increase (blue). Eventually, the spectrum softens as a result
of the soft X-ray accretion disk increasing in luminosity and the
power law itself becoming softer. In the \textit{intermediate state},
both an accretion disk and a power-law are prominent in the spectrum
(green). The source enters the soft state when the disk begins
to dominate the X-ray flux (red). We see 
that the source undergoes a hysteresis loop in which the luminosity is
lower on the return to quiescence than during the rise to
outburst. For a more detailed review, see e.g. Belloni (2010); Belloni
et al (2005).

The spectral transitions can be explained by the \textit{truncated disk
model} (Ichimaru 1977; Done et al 2007). In the hard state, the
geometrically thin, optically thick accretion disk truncates at a
radius larger than the innermost stable circular orbit (ISCO). The
disk is therefore very weak in the spectrum and also comparatively
cool. Inside of the truncation radius is a large scale height, hot
accretion flow. Compton up-scattering of cool disk photons by hot
electrons in this \textit{hot inner flow} creates the power-law
spectrum, which dominates. As the mass accretion rate increases to
make the source brighter, the truncation radius moves in. This means
that the inner flow is cooled by a greater luminosity of comparatively
cool disk photons, leading the power-law to soften. This also
naturally leads to the disk becoming more prominent in the
spectrum. Alternative interpretations of the spectrum are prominent in
the literature, in particular many authors suggest that the power-law
spectrum originates in the base of a jet rather than an inner hot flow
(Markoff, Nowak, \& Wilms; Miller 2007). However, in this review I
focus on the truncated disk model, which has been considered far
more extensively than any other in the context of rapid X-ray
variability (see e.g. Gilfanov 2010).

Variability on shorter timescales is best characterised using Fourier
transforms. The power spectrum\footnote{the modulus squared of the
  Fourier transform} of the X-ray flux evolves just as dramatically as
the X-ray spectrum during an outburst. Examples representative of each
of the three states are shown at the top of Fig \ref{fig:HIDschem} for
the source GX339-4. In the hard state there is a large amplitude of
variability on a broad range of time scales, evidenced by the high
normalisation and broad shape of the power spectrum (right). We can
see low and high frequency breaks at $\sim 0.05$ Hz and $\sim 5$ Hz
respectively where the power drops-off from `flat-top' noise. As the
spectrum evolves to the intermediate state, the low frequency break
moves to higher frequencies and the total variability reduces
(middle). The high frequency noise ($\gtrsim 10$ Hz), in contrast,
stays roughly constant (Gierli{\'n}ski, Niko{\l}ajuk \& Czerny
2008). We can also see narrow peaks in the power spectra. These are
quasi-periodic oscillations (QPOs). Since they pick out a
characteristic time scale in the system, they are very
interesting. However, I will focus on the broad band noise here and
refer the interested reader to Sara Motta's contribution to this
volume. As the spectrum continues to evolve (and the truncation radius
moves further in according to the truncated disk model), the low
frequency break moves to still higher frequencies and the total
variability amplitude continues to reduce until, in the soft state,
there is barely any variability at all (left). See e.g. van der Klis
(2006) for a more detailed review.

Here, I review a paradigm that has emerged over the last $\sim 15$
years to explain the properties of the rapid aperiodic X-ray
variability in the context of the truncated disk model. It is
increasingly thought that this variability is due to fluctuations in
the mass accretion rate. In Section \ref{sec:theory} I review the
theoretical motivation for considering propagating accretion rate
fluctuations. In Section \ref{sec:obs} I review the main observational
evidence for the model and in Section \ref{sec:discussion} I discuss
the remaining challenges to address in future.

\section{Theoretical background}
\label{sec:theory}

The propagation of fluctuations in an accretion flow depends on
viscosity and differential rotation. As gas spirals towards the black
hole, the infall velocity is likely small compared with the orbital
velocity, $v_R << v_\phi$, which is thus Keplerian to a good
approximation, $v_\phi = \sqrt{GM/R}$. For gas at radius $R$ to drift
to radius $R-dR$, it must lose angular momentum (i.e. be slowed down)
through viscous shear from the differentially rotating stream of gas
inside of it. The force per unit surface area (the stress) felt by the
stream at $R$ is $-\eta R \Omega'$, where $\Omega'$ is the radial
derivative of the orbital angular velocity ($\Omega=v_\phi/R$) and
$\eta$ is the \textit{dynamic viscosity} (Frank, King \& Raine 2002;
hereafter FKR02). The \textit{kinematic viscosity} is defined as
$\nu=\eta/\rho$, where $\rho$ is the gas density. This arises in a
fluid through some combination of molecular transport (i.e. thermal
motion of molecules in the fluid) and hydrodynamic
turbulence. Accretion flows are thought to be dominated by the latter,
since the former cannot generate a high enough viscosity (FKR02). We
can understand the kinematic viscosity by expressing it as $\nu \sim
\lambda \tilde{v}$, where $\lambda$ and $\tilde{v}$ are the
characteristic size and speed of the turbulence. In the absence of a
deep understanding of the physics underpinning the viscosity, Shakura
\& Sunyaev (1973) famously used the parameterisation, $\nu = \alpha
c_s H$, where $c_s$ is the sound speed, $H$ is the height of the disk/flow
and $\alpha$ is the dimensionless viscosity parameter. This comes from
assuming that the turbulent motion is probably not supersonic, so
$\tilde{v} \lesssim c_s$, and the size scale of the turbulent eddies
is not larger than the disk height $\lambda \lesssim H$. This implies
that $\alpha \lesssim 1$ is reasonable, although there is no \textit{a
  priori} reason to believe that $\alpha$ is independent of
radius. Over the past few decades, the magnetorotational instability
(MRI: Balbus \& Hawley 1991) has emerged as a strong candidate
mechanism to provide the required turbulence. This is effectively
tangling up of magnetic field lines in the differentially rotating gas.

\begin{figure}
\centering
 \includegraphics[height= 90mm,width=70mm,trim=0.0cm 0.0cm 0.0cm
0.0cm,clip=true]{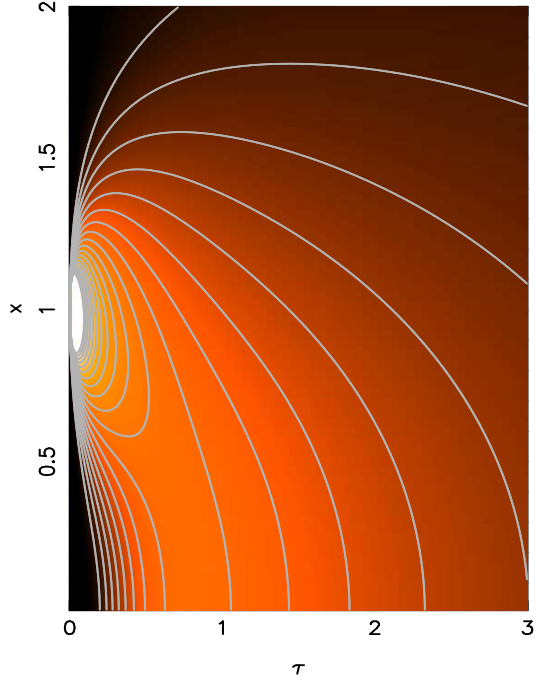}
 \caption{
Representation of the Greens function from equation
\ref{eqn:greens}. The dimensionless time and radius variables are
respectively $\tau$ and $x$. A $\delta-$function perturbation in the
surface density originates at $\tau=0$, $x=1$. Lighter shades
represent higher surface density, and so the perturbation initially
shows up as a narrow bright patch on the plot (this saturates the
colour scheme and so the perturbation appears to have finite
width). The perturbation spreads out over time and moves towards
smaller $x$. This is made clearer by the grey contours of constant
surface density. }
\label{fig:Greens}
\end{figure}

The evolution of perturbations in the accretion disk can be described
by the \textit{diffusion equation} (Lynden-Bell \& Pringle
1974). Assuming circular, Keplerian orbits and applying mass and
angular momentum conservation, this takes the form
\begin{equation}
\frac{\partial \Sigma}{\partial t} = \frac{3}{R}
\frac{\partial}{\partial R} \left\{ R^{1/2} \frac{\partial}{\partial
    R}\left[ \nu\Sigma R \right] \right\},
\label{eqn:diff}
\end{equation}
where $\Sigma = \rho H$ is the \textit{surface density}. In principle,
this equation is valid both for a classic geometrically thin accretion
disk and also the large scale height inner hot flow thought to be present in the
hard state -- but only if the assumptions of Keplerian orbits and
negligible vertical gradients hold. Although both of these assumptions
may break down for a particularly large scale height $H/R$, they will
approximately hold for reasonable parameters (e.g. Fragile et al
2007). Now let us explore the properties of the diffusion equation by
introducing a $\delta$-function perturbation in the surface density at
radius $R=R_0$ and time $t=0$. Here I explore the illustrative example
set out in FKR02 using $\nu=$constant, although note that more
sophisticated treatments of the viscosity are possible (Lynden-Bell \&
Pringle 1974; Lyubarskii 1997; Kotov et al 2001; Tanaka 2011). In this
case, the surface density of an initially thin ring with mass $m$
evolves in time as
\begin{equation}
\Sigma(x,\tau) = \frac{m}{\pi R_0^2} \tau^{-1} x^{-1/4} \exp\left\{ -
  \frac{(1+x^2)}{\tau} \right\} I_{1/4}(2x/\tau),
\label{eqn:greens}
\end{equation}
where $I_{1/4}$ is a modified Bessel function and radius and time are
expressed as the dimensionless variables $x=R/R_0$ and $\tau=12 \nu t
R_0^{-2}$. The contours and shading in Fig \ref{fig:Greens}
illustrate equation \ref{eqn:greens}. At time $\tau=0$,
the perturbation in surface density is a narrow ring at $x=1$. With
time, this ring spreads out and moves towards the black hole.

In a limit in which equation \ref{eqn:diff} is linear, equation
\ref{eqn:greens} is a Greens function, $g(R,t)$ (with dimensionless
variables replaced by physical ones). Thus, if we were to introduce
fluctuations at radius $R=R_0$ with general form $d(t)$, the surface
density will evolve in time as $\Sigma(R,t) = d(t) \ast g(R,t)$, where
$\ast$ denotes a convolution. The Fourier transform of the surface
density is therefore $\Sigma(R,f) = D(f) G(R,f)$. If $d(t)$ is white
noise, this reduces further to $\Sigma(R,f) \propto G(R,f)$. From mass
conservation, $\dot{M}(R,f) \propto \Sigma(R,f)  \propto G(R,f)$ and
so the Greens function tells us all we need to know about the nature
of mass accretion rate fluctuations in an accretion flow.

As noted by Churazov et al (2001), we can simplify this in the limit
of $R<<R_0$ to gain insight (i.e. taking a horizontal slice across Fig
\ref{fig:Greens} at small $x$). In this case\footnote{since
  $I_{1/4}(z) \propto \sim z^{1/4}$ for small $z$}, equation
\ref{eqn:greens} reduces approximately to $\Sigma \propto \tau^{5/4}
\exp[-1/\tau]$. We see that a $\delta-$function spike in the surface
density at $R=R_0$ causes a much slower rise at $R<<R_0$, followed by
an exponential decay with a characteristic decay timescale of
$\tau=1$. This characteristic timescale is called the viscous
timescale and it follows from the definition of $\tau$ that
\begin{equation}
t_{visc}(R) = \frac{R^2}{12\nu}.
\end{equation}
Fluctuations on timescales shorter than the viscous timescale, or in
other words faster than the viscous frequency $f_{\rm visc}(R) = 1 /
t_{\rm visc}(R)$, are therefore strongly damped by the combination of
viscosity and differential rotation. The Fourier transform of a damped
exponential is a zero-centred Lorentzian, and therefore the power
spectrum of mass accretion rate fluctuations in the accretion disk is
approximately
\begin{equation}
|\dot{M}(R,f)|^2 \propto \frac{1}{1+[f/f_{\rm visc}(R)]^2}.
\label{eqn:lore}
\end{equation}
Note that this equation assumes that Equation \ref{eqn:greens}
approximately takes the form of a damped exponential even for locally
produced fluctuation (i.e. $R=R_0$). Considering that the very
existence of the viscosity in the disk hinges on the presence of
turbulence, it is fair to assume that mass accretion rate fluctuations
are stirred up throughout the accretion flow by the same process that
generated the viscosity. Assuming that the generated fluctuations have
the nature of white noise, the power spectrum of the mass accretion
rate everywhere in the accretion flow is given by equation
\ref{eqn:lore} (Arevalo \& Uttley 2006; Ingram \& Done 2011;
2012). The viscous timescale also sets the timescale in which a
perturbation is accreted. The infall velocity of accreting material is
therefore $v_R(R) \sim R f_{\rm visc}(R)$. This gives us the frame
work of the propagating fluctuations model: perturbations propagate
inwards at a speed set by viscosity and are also damped by viscosity.

So what is the viscous frequency? Using Shakura \& Sunyaev's
$\alpha-$prescription one more time and setting the sound speed to
$c_s = (H/R) v_\phi$ yields
\begin{equation}
f_{visc}(R) \approx 2 \alpha (H/R)^2 f_\phi(R),
\label{eqn:alphavisc}
\end{equation}
where $f_\phi(R)=\Omega(R)/(2\pi)$ is the orbital frequency. From
equation \ref{eqn:alphavisc}, we see that rapid variability is allowed
for a large scale heigh flow, as is thought to be present in the hard
state, but is damped for the thin disk (small $H/R$) present in the
soft state. This is consistent with observations of the variable hard
state and stable soft state (see Fig \ref{fig:HIDschem}). We also
expect the amplitude of variability generated by turbulence to
increase with $H/R$ (Ingram 2012). Imagine turbulent eddies, with
characteristic length scale $\sim H$, generating variability in the
accretion rate. It is possible to fit $N \sim 2\pi /H$ such eddies in
a ring of the accretion flow at radius $R$. If these eddies are all
independent of one another and all generate variability with roughly
the same mean $\mu$ and variance $\sigma^2$, then the entire ring will
have a mean $\mu_{\rm tot} = N \mu$ and variance $\sigma^2_{\rm tot} =
N \sigma^2$. The fractional rms generated by the ring is therefore
$\sigma_{\rm tot} / \mu_{\rm tot} \propto (H/R)^{1/2}
(\sigma/\mu)$. It is therefore expected that a thin disk will be
observed to be more stable than a thick disk.

\section{Observations}
\label{sec:obs}

Many of the rapid X-ray variability properties of black hole binaries
are naturally explained by the propagating mass accretion rate
fluctuations model. In this Section, I cover the most prominent
examples.

\subsection{Power spectrum}

The stable disk / variable inner flow paradigm (e.g. Revnivtsev,
Gilfanov \& Churazov 1999; Churazov et al 2001) predicts that the low
frequency break in the power spectrum is simply set by the viscous
frequency at the outer edge of the inner hot flow. This frequency
increases as the truncation radius moves in during spectral evolution,
naturally explaining the observed increase in the low frequency break
(Wijnands \& van der Klis 1999). The \textsc{xspec} model
\textsc{propfluc} (Ingram \& Done 2011; 2012; Ingram \& van der Klis
2013) assumes a variable inner flow with some inner and outer radius,
which are model parameters. The viscous frequency (or rather the
surface density: the two are linked by mass conservation) as a
function of $R$ is  parameterised. Originally, the process of
stochastic fluctuations propagating towards the black hole could only
be calculated using Monte Carlo simulations. Ingram \& van der Klis
(2013) developed a formalism to perform exactly the same calculation
analytically without making any further assumptions. \textsc{propfluc}
additionally assumes that the QPO results from Lense-Thirring
precession of the inner flow (Ingram, Done \& Fragile 2009). There are
now a few studies in the literature that fit the \textsc{propfluc}
model to observed power spectra (Ingram \& Done 2012; Rapisarda,
Ingram \& van der Klis 2014). See Stefano Rapisarda's contribution to
this volume for a more comprehensive summary of the \textsc{propfluc}
model.

\subsection{Linear RMS-flux relation}

A linear RMS-flux relation was first discovered for the case of Cygnus
X-1 by Uttley \& McHardy (2001). Roughly speaking, this can be
measured by chopping a long light curve up into short segments and
calculating the mean flux for each segment and the standard deviation
(absolute rms) around each of these flux measurements. Plotting the
absolute rms measurements against the flux measurements, after binning
in flux, reveals a striking linear relation. This relation holds for a
large range of timescales -- whether we use $2$s segments or $20$s
segments, we still see a linear rms flux relation for Cygnus X-1. The
linear rms-flux relation seems to be a ubiquitous property of variable
accreting objects (Uttley 2004; Heil \& Vaughan 2010; Heil, Vaughan \&
Uttley 2012). This property of the variability tells us that different
timescales are correlated. In the picture laid out in the previous
section in which different timescales are produced in different
regions, the linear rms-flux relation tells us that these regions must
be \textit{causally connected}. Uttley, McHardy \& Vaughan (2005)
showed that this property rules out previously popular shot noise
models, in which the variability is modelled as a sum of unrelated
flares drawn from some distribution.

\begin{figure}
\centering
 \includegraphics[height= 70mm,width=80mm,trim=0.5cm 0.0cm 0.0cm
0.0cm,clip=true]{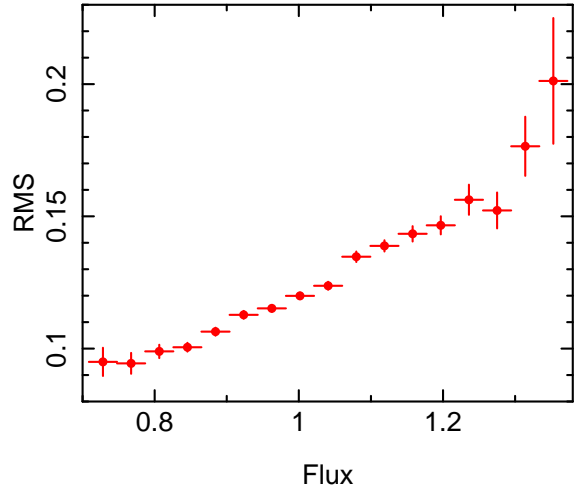}
\vspace{-5mm}
 \caption{RMS-flux relation for a simulation using the model
   \textsc{propfluc}. The outer and inner radii of the variable hot
   inner flow are assumed to be $50 R_g$ and $4.25 R_g$
   respectively. A $8192$s light curve was generated using a Monte
   Carlo simulation and split up into $4096$ $2$s segments. The
   relation is linear, consistent with observations.}
\label{fig:rmsflux}
\vspace{-5mm}
\end{figure}

The propagating fluctuations model naturally predicts a linear
rms-flux relation. This happens because the fluctuations produced at
all radii propagate inwards. So, a fluctuation generated far from the
black hole propagates inwards to modulate the faster variability
produced closer to the black hole. Since this modulation is a
multiplicative process, this results in intervals of low(high) flux having a
low(high) variability amplitude\footnote{i.e. When the flux from the
  outer region is low, the variability amplitude from the inner region
is multiplied by a small number and when the flux from the outer
region is high, the amplitude from the inner region is multiplied by a
large number.}. Fig \ref{fig:rmsflux} shows the an rms-flux relation
calculated with \textsc{propfluc}. Here, I generated a long light
curve ($8192$s) and split it into $2$s segments (it is still not
possible to calculate the rms-flux relation predicted by the model
analytically and so this plot required a Monte Carlo simulation). We
see that it is indeed linear as expected.

\subsection{Frequency-resolved spectroscopy}

Another basic ingredient of the propagating fluctuations paradigm is that
harder radiation is generally emitted closer to the black hole. This
is simple to justify theoretically, since the gravitational energy
loss experienced by accreting material increases closer to the black
hole. For a thermalised accretion disk, the temperature goes as
$T\propto R^{-3/4}$. For the inner flow, the outer region is cooled
by a greater luminosity of comparatively cool disk photons than the
inner region (which is further from the disk). We therefore expect the
spectrum emitted by the inner flow to be softer at larger
radius.

Revnivtsev et al (1999) used frequency-resolved spectroscopy
to reinforce this picture. The frequency-resolved
spectrum for the frequency range $f_1$ to $f_2$ is simply the absolute
variability amplitude in this frequency range as a function of 
energy. Revnivtsev et al (1999) showed that the spectrum of the slow
variability in Cygnus X-1 in the hard state is a hard power law, and
the spectrum of the faster variability is a softer power law. This
fits in nicely with the picture that the fast variability and hard
radiation originate from small $R$ and the slow variability and soft
radiation originate from large $R$.

\subsection{Time lags}

\begin{figure}
\centering
 \includegraphics[height= 70mm,width=80mm,trim=0.5cm 0.0cm 0.0cm
0.0cm,clip=true]{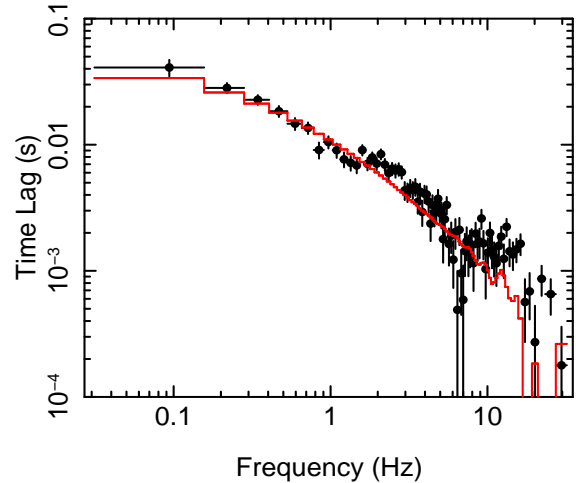}
\vspace{-5mm}
 \caption{
Time lag of the $\sim 5.6-16.6$ keV band with respect to the $\sim
2-5.6$ keV band as a function of Fourier frequency. The black points
are for an \textit{RXTE} observation of Cygnus X-1 (obs ID
10238-01-08-00). The red line is calculated from the model
\textsc{propfluc}, assuming that a harder spectrum is emitted at
smaller radii in the inner hot flow.}
\label{fig:Lag}
\vspace{-5mm}
\end{figure}

Since fluctuations take time to propagate towards the black hole, the
model predicts photons emitted at small $R$ to lag photons emitted at
large $R$. Since the spectrum is harder closer to the black hole, the
photons emitted at small $R$ are, on average, harder than those
emitted at large $R$ and therefore hard photons lag soft photons. We
cannot directly observe which radius photons are emitted from, but we
can isolate different Fourier frequencies. The model predicts that the
time lag between hard and soft photons is longer for slow variability
(low Fourier frequency) than for fast variability (high
Fourier frequency). This is because slow variability can be generated
anywhere in the flow but fast variability can only be generated at
small $R$. Therefore, on average, fast fluctuations have not
propagated as far between emitting soft and hard photons. Figure
\ref{fig:Lag} shows the time lag between two broad energy bands as a
function of Fourier frequency for an observation of Cygnus X-1 (black
points). Here, a positive lag means that hard photons lag soft. We see
that hard photons do indeed lag soft photons, and the amplitude of the
lag reduces with Fourier frequency. The red line models the lag using
\textsc{propfluc} (Ingram \& van der Klis 2013 found an analytic
expression for the time lags in addition to the power spectrum). The
propagation time between a given pair of radii is set by the viscous
frequency (since $v_R(R) \sim R f_{\rm fisc}(R)$), and the radial
dependence of spectral hardness is parameterised by assuming a
different (power-law) radial emissivity profile for each energy
band.

The time lag is also observed to depend on the energy bands
considered. The greater the separation between the energy bands, the
longer the lag (e.g. Kotov et al 2001). This makes intuitive sense in
the propagating fluctuations picture, since the softest photons are
predominantly emitted at the largest $R$ and so fluctuations from here
must propagate further to reach the smallest $R$ where the hardest
photons are emitted.

\section{Discussion \& Conclusions}
\label{sec:discussion}

The propagating fluctuations model can successfully explain many
observational properties of black hole binaries, but there are a
number of improvements that can be made and challenges to be
overcome. Perhaps most obvious is the details of the observed power
spectra and lag spectra. In Fig. \ref{fig:HIDschem}, we see that the
broadband noise in the observed power spectrum (presented in units of
frequency $\times$ power) looks roughly like flat top noise, with high
and low frequency breaks. \textsc{propfluc} produces \textit{exactly}
this shape, since it assumes that each radius of the flow generates
fluctuations of the same amplitude. For some observations, however, it
is clear that the broad band noise has more structure than this, with
lumps and bumps often modelled as individual Lorentzian
components. This structure is also in the lag
spectra. Fig. \ref{fig:Lag} shows lumps in the data that are not
reproduced by the model. Features tend to be present at the same
frequency for both the power and lag spectra of a given
observation (Grinberg et al 2014). For example, the power spectrum
corresponding to the lag 
spectrum shown in Fig. \ref{fig:Lag} also has a hump peaking at $\sim
2-3$ Hz. It is in principle possible to reproduce these features with
\textsc{propfluc} by simply allowing the variability generated in the
flow to depend on radius. Ingram \& Done (2012) demonstrated that a
`double hump' power spectrum can be produced by introducing enhanced
variability at one particular radius (Figure 6 therein). This also
impacts on the  predicted lag spectrum, so even an \textit{ad hoc}
introduction of variability at a particular radius, creates features
in the power \textit{and} lag spectrum, that must be both consistent
with data. In other words, a tweak that allows the model to fit the
power spectrum also inevitably changes the predicted lag spectrum.

We of course should also have a good reason to introduce extra
variability at a given radius. Henisey et al (2012) reported on
enhanced variability at $\sim 7R_g$ in numerical simulations of a
large scale height accretion flow caused by the misalignment of the
flow spin axis with that of the black hole. The frame dragging effect
(the twisting up of the surrounding spacetime by the spinning black
hole) means that the black hole itself exerts a torque on the
accretion flow. Fragile et al (2007) showed that this introduces
`plunging streams' of material at $\sim 7R_g$ in simulations in which
the black hole spin axis is misaligned with the flow spin axis. In
these plunging streams, material falls towards the black hole
relatively quickly in constrast to the slow viscous drift at larger
radii. Henisey et al (2012) showed that there is actually enhanced
variability in the vicinity of these streams. Enhanced variability
could alternatively / additionally come from the disk. When the
standard picture was of stable disk and variable inner flow, Wilkinson
\& Uttley (2009) found that the disk was actually variable on
timescales $\gtrsim 1$s in hard state observations of GX339-4 and
SWIFT J1753.5−0127. Therefore, bumps in the power spectra at low
frequency could potentially come from variability in the disk rather
than the flow (see Stefano Rapisarda's contribution to this
volume). This makes a prediction in terms of the lag spectrum, since
fluctuations from the disk take time to propagate to the hard emitting
region, which can be tested directly against data (Rapisarda et al in
prep).

A significant improvement to the model would be to use a more
sophisticated Green's function to model the damping and propagation of
fluctuations, rather than using the current Lorentzian treatment
(i.e. equation \ref{eqn:lore}) with propagation occurring at the
viscous infall speed. It is relatively easy to make this change if the
model calculation uses a Monte Carlo simulation. It is, however, far
more difficult to adapt the analytic formalism of Ingram \& van der
Klis (2013) for the case of a general Greens function, which is
required for the model to realistically be tested against data.

Another important physical effect that has not been mentioned thus
far is reflection of hard X-ray photons from the outer disk. Reflected
photons contribute a relatively small fraction of the observed flux
for most observations of black hole binaries but \textit{certainly}
not a negligible fraction. Also, reflection provides a powerfull
diagnostic of the accretion flow, since the narrow iron line emitted
due to reflection is distorted in a characteristic way by the orbital
motion of disk material. This effect was considered in the context of
propagating mass accretion rate fluctuations by Kotov et al (2001) and
should also be included in future versions of the \textsc{propfluc}
model.

In conclusion, the general propagating fluctuations picture is very
well motivated theoretically and the data provide strong evidence
to back this up. In particular the rms-flux relation is a natural
consequence of propagating fluctuations but very difficult to contrive
in alternative scenarios. The details of the model, however, are still
uncertain. Systematic, quantitative testing of increasingly
sophisticated versions of the model against the huge archive of public
data provided by \textit{RXTE} will address this going forward.

\acknowledgements
I acknowledge support from the Netherlands Organisation for
Scientific Research (NWO) Veni Fellowship. This research has
made use of data obtained through the High Energy Astrophysics
Science Archive Research Center Online Service, provided by the
NASA/Goddard Space Flight Center.

\end{document}